\RequirePackage{fix-cm}

\documentclass[twocolumn]{svjour3}          
\smartqed                                   

\usepackage{graphicx}
\usepackage{mathptmx}
\usepackage{amsmath,amssymb}
\usepackage[numbers,comma,sort&compress]{natbib}
\usepackage[T1]{fontenc}
\usepackage{color}
\usepackage[titletoc,title]{appendix}

\DeclareMathAlphabet{\mathcal}{OMS}{cmsy}{m}{n}

\renewcommand{\thesubsection}{\Alph{subsection}}

\begin{document}

\title{Analytical evaluation of relativistic molecular integrals.\\
II. Computational aspect for relativistic molecular auxiliary functions.}

\author{
Ali Ba{\u g}c{\i}
\and
Philip E. Hoggan
\and
Muzaffer Adak
}

\institute{
A. Ba{\u g}c{\i} \at
Department of Physics, Faculty of Arts and Sciences, Pamukkale University 20017 Denizli, Turkey
\\
\email{abagci@pau.edu.tr}
\and
P. E. Hoggan \at
Institute Pascal, UMR 6602 CNRS, University Blaise Pascal, 24 avenue des Landais BP 80026, 63177 Aubiere Cedex, France
\and
M. Adak \at
Department of Physics, Faculty of Arts and Sciences, Pamukkale University 20017 Denizli, Turkey
}

\date{Received: date / Accepted: date}
\maketitle

\begin{abstract}
The Slater-type orbital basis with non-integer principal quantum numbers is a physically and mathematically motivated choice for molecular electronic structure calculations in both non-relativistic and relativistic theory. The non-analyticity of these orbitals at $r=0$, however, requires analytical relations for multi-center integrals to be derived. This is nearly insurmountable. Previous papers by present authors eliminated this difficulty. Highly accurate results can be achieved by the procedure described in these papers, which place no restrictions on quantum numbers in all ranges of orbital parameters. The purpose of this work is to investigate computational aspects of the formulae given in the previous paper. It is to present a method which helps to increase computational efficiency. In terms of the processing time, evaluation of integrals over Slater-type orbitals with non-integer principal quantum numbers are competitive with those over Slater-type orbitals with integer principal quantum numbers.

\end{abstract}

\section{Introduction}\label{introduction}
In the first paper \cite{1_Bagci_2018} of aforementioned series, history and importance of usage the auxiliary function method was summarized. Applications in molecular calculations were briefly given. A discussion was made on the relativistic molecular auxiliary functions introduced previously \cite{2_Bagci_2015}. They are used when the principal quantum numbers in a Slater-type orbital basis (STO) \cite{3_Slater_1930}  are free from any restriction \cite{4_Parr_1957},
\begin{equation}\label{eq:intro_1}
\chi\left(\zeta, \vec{r} \right)
=\frac{\left(2\zeta\right)^{n+1/2}}{\sqrt{\Gamma\left(2n+1 \right)}}
r^{n-1}e^{-\zeta r}Y_{lm}\left(\theta, \varphi \right),
\end{equation}
The $Y_{lm}$ are complex or real spherical harmonics $\left( Y^{*}_{lm}=Y_{l-m} \right.;$ $ \left.S_{lm}\equiv Y_{lm} \right)$ \cite{5_Condon_1935}. The STO basis with non-integer pribcipal quantum numbers provides extra flexibility for closer variational description of trial wavefunction \cite{4_Parr_1957, 6_Geller_1962, 7_Allouche_1974, 8_Allouche_1976, 9_Koga_1997, 10_Koga_1997, 11_Koga_1997, 12_Koga_1998, 13_Koga_2000, 14_Guseinov_2009, 15_Guseinov_2012} in the linear combination of atomic orbital method  \cite{16_Roothaan_1951}. They also lead to use of a Slater-type spinor basis \cite{17_Grant_2007, 18_Bagci_2016} in algebraic solution of the four-component Dirac equation \cite{19_Quiney_2002, 20_Belpassi_2008} due to the so-called kinetic balance condition \cite{21_Lee_1982, 22_Richard_1984, 23_Esteban_1999, 24_Grant_2010, 25_Kutzelnigg_2012, 26_Lewin_2014}. The matrix elements arising in a generalized eigenvalue equation are evaluated through prolate spheroidal coordinates and expressed in terms of relativistic molecular auxiliary functions. A method to analytically evaluate these auxiliary functions was obtained via convergent series representation of incomplete beta functions. They were derived according to a criterion \cite{1_Bagci_2018}. Symmetry properties arise from evaluating two-center two-electron integrals via the two-range addition theorem and give this result. The incomplete gamma functions are thereby eliminated from auxiliary functions via up-ward or down-ward recurrence relations. The relations obtained for two-center two-electron integrals are compact, expressed using overlap integrals.

The main goal of the present study is to open the lock to usability of the analytical approach. Thus, derivation of the relativistic molecular auxiliary functions is re-visited. Regarding analytical evaluation of overlap integrals, differences and similarities between using integer and non-integer principal quantum numbers in STOs are investigated. The analytical expressions we previously obtained are used to calculate starting values. Computationally efficient recurrence relations are derived accordingly. A simple computational scheme on how to use the presented formulae is described.

Range-separated functionals \cite{27_Iikura_2001} which use STOs and Yukawa-like potentials \cite{28_Yukawa_1935} for the attenuated electron-electron interaction in density functional theory have focussed interest in recent years \cite{29_Bouferguene_2005, 30_Angyan_2006, 31_Seth_2012, 32_Rico_2013}. Procedure described for evaluation of two-center integrals over STOs and Coulomb potential is used for equivalent integrals with the Yukawa-like potentials. The relativistic molecular auxiliary functions are available to be used in evaluation of these integrals for any potentials. In addition, the STOs to be used are free from any restriction. 

\section{Origin of the relativistic molecular auxiliary functions}\label{origin}
Following the procedure given in \cite{33_Weatherford_2006} the expression of two-center two-electron Coulomb energy associated with a charge density $\rho\left(\vec{r} \right)$,
\begin{equation}\label{eq:origin_2}
E=\int\int G\left(\vec{r}_{1},\vec{r}_{2} \right)\rho\left(\vec{r}_{a_{1}} \right)\rho\left(\vec{r}_{b_{2}} \right)dV_{1}dV_{2}
\end{equation}
where, $G\left(\vec{r}_{1},\vec{r}_{2} \right)$ is the Green's function for the Laplace equation.  The Coulomb operator is transformed into a kinetic-energy-like integral using Poisson's equation for the density due to electron indicated by 2,
\begin{equation}\label{eq:origin_3}
\bigtriangledown^{2}_{\vec{r}_{b_{2}}}V\left(\vec{r}_{b_{2}} \right)
=-4\pi \rho\left(\vec{r}_{b_{2}} \right)
\end{equation}
and single-center potential,
\begin{equation}\label{eq:origin_4}
V\left(\vec{r}_{a_{2}} \right)
=\int \frac{\rho \left(\vec{r}_{a_{1}} \right)}{r_{12}}dV_{1}
\end{equation}
as,
\begin{multline} \label{eq:origin_5}
E=-{\frac{1}{4\pi }}
\int{V\left(\vec{r}_{a_{2}}\right)\bigtriangledown_{\vec{r}_{b_{2}}}^{2}V^{\text{*}}
\left(\vec{r}_{b_{2}}\right)}\mathit{dV}_{2}\\
=\int V\left(\vec{r}_{a_{2}}\right)\rho
\left(\vec{r}_{b_{2}}\right)\mathit{dV}_{2}.
\end{multline}
Here, $\mathit{dV}=r^{2}\sin \left(\theta \right)d\theta d\phi$, the integration domain is $\left[0,\infty \right)\times\left[0,\pi \right]\times\left[0,2\pi\right]$. Note that these expressions are symmetric with respect to exchange in subscripts $a,b$.\\
For normalized non-integer Slater-type orbital (NSTO) the one-center potential are expressed in terms of radial functions as \cite{2_Bagci_2015},
\begin{align} \label{eq:origin_6}
V(r_{a_{2}})=\sum
_{L_{1}M_{1}}{F_{N_{1}}^{L_{1}}\left(x_{1},\vec{r}_{a_{2}}\right)\mathcal{C}_{L_{1}M_{1}}}.
\end{align}
where,
\begin{multline} \label{eq:origin_7}
F_{N_{1}}^{L_{1}}\left(x_{1},\vec{r}_{a_{2}}\right)\\
=\mathcal{N}_{n_{1}n_{1}'}\left(1,t_{1}\right)\left(2\bar{\zeta
}_{1}\right)f_{N_{1}}^{L_{1}}\left(x_{1},r_{a_{2}}\right)Y_{L_{1}M_{1}}^{\text{*}}(\theta
_{a_{2}}\phi _{a_{2}})
\end{multline}
with,
\begin{multline} \label{eq:origin_8}
f_{N_{1}}^{L_{1}}\left(x_{1},r_{a_{2}}\right)=
\Gamma\left(N_{1}+L_{1}+1\right)\frac{1}{x_{1}^{L_{1}+1}}\biggl\lbrace P[N_{1}+L_{1}+1,x_{1}]\biggl. \\ \biggl. +\frac{x_{1}^{2L_{1}+1}}{(N_{1}-L_{1})_{2L_{1}+1}}Q[N_{1}-L_{1},x_{1}] \biggr\rbrace,
\end{multline}
and,
\begin{multline} \label{eq:origin_9}
\mathcal{C}_{L_{1}M_{1}}
=\left(\frac{4\pi
}{2L_{1}+1}\right)^{1/2}C^{L_{1}M_{1}}(l_{1}m_{1},l_{1}'m_{1}')A_{m_{1}m_{1}'}^{M_{1}},
\end{multline}
are the generalized Gaunt coefficients, see \cite{34_Guseinov_1970, 35_Guseinov_1995} for the definition of $A^{M}$ coefficients. The normalization coefficients are determined by,
\begin{align} \label{eq:origin_10}
\mathcal{N}\left(p,\tau \right)=\frac{[p+t]^{n+1/2}[p-t]^{n'+1/2}}{\sqrt{\Gamma
[2n+1]\Gamma [2n'+1]}}
\end{align}
with, \\
$x=2\bar{\zeta }r$, $\bar{\zeta}=\frac{1}{2}\left(\zeta +\zeta' \right)$, $p=\frac{R}{2}\left(\zeta+\zeta' \right)$, $t=\frac{\zeta-\zeta'}{\zeta+\zeta'}$, $N=n+n'$, $\left\lbrace\zeta, \zeta' \right\rbrace$ are orbital parameters.\\
$P[\alpha ,z]$,  $Q[\alpha ,z]$ are the normalized incomplete gamma and its complement  \cite{36_Abramowitz_1972, 37_Temme_1994},
\begin{equation}\label{eq:origin_11}
P\left[\alpha, z \right]
=\frac{\gamma\left(\alpha,z\right)}{\Gamma\left(\alpha\right)},
\hspace{5mm}
Q\left[\alpha, z \right]
=\frac{\Gamma\left(\alpha,z\right)}{\Gamma\left(\alpha\right)}.
\end{equation}
By definition, they satisfy.
\begin{equation}\label{eq:origin_12}
P+Q=1.
\end{equation}
$\gamma(a,z)$ and $\Gamma(a,z)$ are the incomplete gamma functions,
\begin{equation}\label{eq:origin_13}
\gamma\left(\alpha, z\right)
=\int_{0}^{z} \tau^{\alpha-1}e^{-\tau}d\tau,
\hspace{5mm}
\Gamma\left(\alpha, z\right)
=\int_{z}^{\infty} \tau^{\alpha-1}e^{-\tau}d\tau.
\end{equation}
$\Gamma(a)$ is the gamma function,
\begin{equation}\label{eq:origin_14}
\Gamma\left(\alpha\right)
=\Gamma\left(\alpha, z\right)+\gamma\left(\alpha, z\right).
\end{equation}
The criterion \cite{1_Bagci_2018} that allows the incomplete gamma functions to be eliminated from the two-center two-electron integrals may now be applied to Eq. (\ref{eq:origin_8}). Using the following up- and down-ward distant recurrence relations for normalized incomplete gamma functions \cite{38_Nist_2018},
\begin{multline}\label{eq:origin_15}
\left\lbrace \begin{array}{cc}
P\left[\alpha, z \right]
\\
Q\left[\alpha, z \right]
\end{array} \right\rbrace
=
\left\lbrace \begin{array}{cc}
P\left[\alpha+n, z \right]+e^{- z}\sum_{s=1}^{n}\frac{\left( z\right)^{\alpha+s-1}}{\Gamma\left(a+s \right)}
\\
Q\left[\alpha+n,z \right]-e^{- z}\sum_{s=1}^{n}\frac{\left( z\right)^{\alpha+s-1}}{\Gamma\left(\alpha+s \right)}
\end{array} \right\rbrace,
\end{multline}
\begin{multline}\label{eq:origin_16}
\left\lbrace \begin{array}{cc}
P\left[\alpha, z \right]
\\
Q\left[\alpha, z \right]
\end{array} \right\rbrace
=
\left\lbrace \begin{array}{cc}
P\left[\alpha-n, z \right]-e^{-z}\sum_{s=1}^{n-1}\frac{\left( z\right)^{\alpha-s-1}}{\Gamma\left(\alpha-s \right)}
\\
Q\left[a-n, z \right]+e^{- z}\sum_{s=1}^{n-1}\frac{\left( z\right)^{\alpha-s-1}}{\Gamma\left(\alpha-s \right)}
\end{array} \right\rbrace,
\end{multline}
both $P$ and $Q$ in Eq.(\ref{eq:origin_8}) are synchronized to $P\left[N_{1}+1, x_{1}\right]$, \newline $Q\left[N_{1}+1, x_{1}\right]$. Making use of Eq. (\ref{eq:origin_12}), the relationship given for Mulliken functions \cite{39_Mulliken_1949},
\begin{equation}\label{eq:origin_17}
A_{\alpha}\left(z \right)=\left(z \right)^{-\alpha-1}\Gamma\left(\alpha+1,z \right),
\end{equation}
the final expression that holds for the incomplete gamma functions obtained as:
\begin{multline}\label{eq:origin_18}
\frac{1}{x_{1}^{L_{1}+1}}\left\lbrace P[N_{1}+1,x_{1}] +\frac{x_{1}^{2L_{1}+1}}{(N_{1}-L_{1})_{2L_{1}+1}}Q[N_{1}+1,x_{1}]\right\rbrace
\\
=\frac{1}{x_{1}^{L_{1}+1}}\left\lbrace
1-\frac{x_{1}^{N_{1}+1}A_{N_{1}}\left(x_{1}\right)}{\Gamma\left(N_{1}+1 \right)}
\left(1+ \frac{x_{1}^{2L_{1}+1}}{(N_{1}-L_{1})_{2L_{1}+1}}\right)
\right\rbrace .
\end{multline}
Such an early operation however not only increases the complexity of relations, since it prevents taking advantage of Eq. (\ref{eq:origin_12}) and leaves no choice but use of the infinite series representation of incomplete gamma functions. It also makes the convergence of the results doubtful. The incomplete gamma functions in the region $0 \leq \alpha < 1 $ are unstable \cite{37_Temme_1994, 40_Gautschi_1999, 41_Chaudhry_2002}. Generation of the incomplete gamma functions by means of recurrence relations for $0 \leq \alpha < 1 $ in an efficient approach and computing the gamma functions without erroneous last digits is still being studied in the
literature \cite{42_Cuyt_2006, 43_Backeljauw_2014, 44_Greengard_2018}. \\
Besides, performing this operation in advance may block analogously generalization of relativistic molecular auxiliary functions \cite{2_Bagci_2015},
\begin{multline} \label{eq:origin_19}
\left\lbrace \begin{array}{cc}
\mathcal{P}^{n_1,q}_{n_{2}n_{3}n_{4}}\left(p_{123} \right)
\\
\mathcal{Q}^{n_1,q}_{n_{2}n_{3}n_{4}}\left(p_{123} \right)
\end{array} \right\rbrace
\\
=\frac{p_{1}^{\sl n_{1}}}{\left({\sl n_{4}}-{\sl n_{1}} \right)_{\sl n_{1}}}
\int_{1}^{\infty}\int_{-1}^{1}{\left(\xi\nu \right)^{q}\left(\xi+\nu \right)^{\sl n_{2}}\left(\xi-\nu \right)^{\sl n_{3}}}\\ \times
\left\lbrace \begin{array}{cc}
P\left[{\sl n_{4}-n_{1}},p_{1}(\xi+\nu) \right]
\\
Q\left[{\sl n_{4}-n_{1}},p_{1}(\xi+\nu) \right]
\end{array} \right\rbrace
e^{p_{2}\xi-p_{3}\nu}d\xi d\nu,
\end{multline}
which are obtained by directly using Eq. (\ref{eq:origin_8}) in the two-center two-electron integrals and making use of the product of two spherical harmonics with the same and different centers in prolate ellipsoidal coordinates $\left(\xi,\nu, \varphi \right)$, with $0 \leq \xi \leq \infty $, $\-1 \leq \nu \leq 1$, $0 \leq \varphi \leq 2\pi$. Here, $\left\lbrace q, n_{1} \right\rbrace \in \mathbb{Z}$, $\left\lbrace n_{2}, n_{3}, n_{4}\right\rbrace \in \mathbb{R}$, $p_{123}=\left\lbrace p_{1}, p_{2}, p_{3}\right\rbrace$ (and in subsequent notation), $p_{1}>0$, $p_{2}>0$, $-p_{2}\leq p_{3} \leq p_{2}$. Taking into account the notation used immediately after Eq. (\ref{eq:origin_10}) , $\left\lbrace p_{1}, p_{2} \right\rbrace = p$, $p_{3}=p t$.\\
The Eq. (\ref{eq:origin_8}) is obtained by expanding Eq. (\ref{eq:origin_4}) using the new set of functions, with the Laplace expansion for Coulomb interaction,
\begin{multline} \label{eq:origin_20}
\frac{1}{r_{12}}
=\sum _{l=0}^{\infty }\sum
_{m=-l}^{l}{\left(\frac{4\pi
}{2l+1}\right)\frac{r_{\text{{\textless}}}^{l}}{r_{\text{{\textgreater}}}^{l+1}}Y_{\mathit{lm}}(\theta
_{1},\varphi _{1})Y_{\mathit{lm}}^{\text{*}}(\theta _{2},\varphi _{2})},
\end{multline}
 $r_{<}$ and $r_{>}$ depend simply on $r_{1}$ and $r_{2}$ through $\vert r_{<} \vert =min\left[r_{1},r_{2} \right]$, $\vert r_{>} \vert =max\left[r_{1},r_{2} \right]$. The Laplace expansion is a two-range series representation of the Coulomb potential $\frac{1}{r_{12}}=\frac{1}{\vert \vec{r}_{1}-\vec{r}_{2} \vert}$, which is not analytic when $r_{1}$ = $r_{2}$. This point is the singularity of the potential. This series expansion is derived by three-dimensional Taylor expansion using the translation operator in general has the form \cite{45_Weniger_2000, 46_Weniger_2002}:
\begin{equation}\label{eq:origin_21}
f\left(r_{<}, r_{>} \right)
=\sum_{s=0}^{\infty}\frac{\left(\vec{r}_{<} \hspace{1mm} . \hspace{1mm} \bigtriangledown_{>} \right)^{s}}{\Gamma[s+1]}f\left(\vec{r}_{>} \right)
=e^{\vec{r}_{<} \hspace{1mm} . \hspace{1mm} \bigtriangledown_{>}}f\left(\vec{r}_{>} \right),
\end{equation}
here, the term $\left( e^{\vec{r}_{<} \hspace{1mm} . \hspace{1mm} \bigtriangledown_{>}} \right)$ given as \cite{47_Santos_1973},
\begin{multline}\label{eq:origin_22}
e^{\vec{r}_{<} \hspace{1mm} . \hspace{1mm} \bigtriangledown_{>}}
=2\pi \sum_{l=0}^{\infty}\sum_{m=-l}^{l}
\mathcal{Y}_{lm}\left(\vec{r}_{<} \right)^{\star}
\mathcal{Y}_{lm}\left(\vec{r}_{>} \right)
\\
\times \sum_{s=0}^{\infty}\frac{\vec{r}_{<}^{2s}\bigtriangledown_{>}^{2s}}{2^{l+2s}s! \left( 1/2 \right)_{l+s+1}},
\end{multline}
with $\mathcal{Y}_{lm}$ are the regular spherical harmonics given by \cite{48_Hobson_1931},
\begin{equation}\label{eq:origin_23}
\mathcal{Y}_{lm}\left(\vec{r} \right)=
r^{l}Y_{lm}\left(\theta, \varphi \right),
\end{equation}
From Eqs. (\ref{eq:origin_21}, \ref{eq:origin_22}) and details of the procedure for constructing an addition theorem \cite{45_Weniger_2000}, we notice that the expression $\left(\xi+\nu \right)$ arising in the Eq. (\ref{eq:origin_19}) can be considered to generate the Coulomb potential. We also pay special attention to the incomplete gamma functions. Extending this, a completely general form of generating function in which the whole set of physical operators can be represented through an argument $f_{ij}^{k}$ has been devised in our previous work. \cite{2_Bagci_2015},
\begin{equation} \label{eq:origin_24}
f_{ij}^{k}=\left(\xi \nu \right)^{k}\left(\xi +\nu \right)^{i}\left(\xi - \nu \right)^{j}.
\end{equation}
They are irreducible representations of elements required to generate the potential. For the Coulomb potential,  the special case when $i=1$, $k=j=0$ $\left( f_{10}^{0}= \left(\xi+\nu \right)\right)$ is appropriate. This means that solution of the Eq. (\ref{eq:origin_5}) for any potential may be represented in terms of Eq. (\ref{eq:origin_19}).\\
The expressions for evaluation of the single-center potential (Eq. (\ref{eq:origin_4}))with Yukawa-like form, use the following two-range formula, derived from Eqs. (\ref{eq:origin_21}, \ref{eq:origin_22}) and give:
\begin{multline}\label{eq:origin_25}
\frac{e^{-\eta r_{12}}}{r_{12}}
=4\pi \sum_{l=0}^{\infty}\sum_{m=-l}^{l}
\left(2l+1 \right) \frac{I_{l+1/2}\left(\eta r_{<} \right)K_{l+1/2}\left(\eta r_{>} \right)}{\sqrt{r_{1}r_{2}}}\\
\times Y_{\mathit{lm}}(\theta
_{1},\varphi _{1})Y_{\mathit{lm}}^{\text{*}}(\theta _{2},\varphi _{2}),
\end{multline}
where, $I_{\alpha+1/2}\left( z \right)$, $K_{\alpha+1/2}\left( z \right)$ are modified Bessel functions of the first- and second-kind (Macdonald functions) \cite{49_Arfken_1985}, respectively, have been obtained in terms of the incomplete gamma functions. This is due to the integral representation of Bessel functions having the following forms (used for single-center potentials):
\begin{equation}\label{eq:origin_26}
\mathrm{I}_{n,l}\left(\zeta,\eta,r \right)
=\int_{0}^{r}\tau^{n+l+1/2}I_{l+1/2}\left(\eta \tau \right)e^{-\zeta \tau}d\tau,
\end{equation}
\begin{equation}\label{eq:origin_27}
\mathrm{K}_{n,l}\left(\zeta,\eta,r \right)
=\int_{r}^{\infty}\tau^{n+l+1/2}K_{l+1/2}\left(\eta \tau \right)e^{-\zeta \tau}d\tau.
\end{equation}
And, the Bessel functions are written in closed form as \cite{50_Gradshteyn_1980},
\begin{multline}\label{eq:origin_28}
I_{\pm \left(l+1/2\right)}\left(z \right)=
\frac{1}{\sqrt{2\pi z}}\Big[
e^{z}\sum_{s=0}^{l}\left(-1 \right)^{s}\frac{\left(l+s \right)!}{s!\left(l-k \right)!}\frac{1}{\left(2z \right)^{s}}
\Big.
\\
\pm \Big.
\left(-1 \right)^{i+1}e^{-z}
\sum_{s=0}^{l}
\frac{\left(l+s \right)!}{s!\left(l-s \right)!}
\frac{1}{\left(2z \right)^{s}}
\Big],
\end{multline}
\begin{equation}\label{eq:origin_29}
K_{l+1/2}\left(z \right)=
\sqrt{\frac{\pi}{2z}}e^{-z}
\sum_{s=0}^{l}\frac{\left(l+s \right)!}{s!\left(l-s\right)!}
\frac{1}{\left(2z \right)^{s}}.
\end{equation}
It is clear from these expressions and Eq. (\ref{eq:origin_13}) that the Eqs. (\ref{eq:origin_26}, \ref{eq:origin_27}) are expressed in terms of incomplete gamma functions. This is, therefore, the only condition needed for two-center two-electron integrals to be easily represented by the relativistic auxiliary functions. They prevent immediate expansion of the incomplete gamma functions in the one-center potentials and make it possible to benefit from the symmetry properties pointed out in the paper.
\section{The two-center overlap integrals}\label{overlap}
By applying the criterion given in \cite{1_Bagci_2018} to the resultant expressions of the two-center two-electron integrals, the relativistic molecular auxiliary functions, for all values of argument $f_{ij}^{k}$, are reduced to overlap-like integrals, which have the following form (in the prolate spheroidal coordinates),
\begin{multline}\label{eq:overlap_30}
\mathcal{G}^{n_{1},q}_{n_{2}n_{3}}\left(p_{123} \right)
=\frac{p_{1}^{n_{1}}}{\Gamma\left(n_{1}+1 \right)}
\\
\times \int_{1}^{\infty}\int_{-1}^{1} \left(\xi\nu\right)^{q}\left(\xi+\nu\right)^{n_{2}}\left(\xi-\nu \right)^{n_{3}}e^{-p_{2}\xi-p_{3}\nu}d\xi d\nu.
\end{multline}
The factor $\left(\xi \nu \right)^{q}$ arises from products of two associated Legendre functions, with different centers \cite{34_Guseinov_1970, 51_Pople_1970},
\begin{multline}\label{eq:overlap_31}
P_{l\lambda }(\cos \theta _{a})
P_{l'\lambda}(\cos \theta _{b})\\
=\sum _{\sigma =-\lambda }^{l}
\sum_{\sigma' =\lambda }^{l'}
\sum _{q=0}^{\sigma +\sigma' }
{g_{\sigma \sigma'}^{q}(l\lambda ,l'\lambda )}\\
\times{\left[\frac{\left( \xi \nu \right)^{q}}
{(\xi +\nu)^{\sigma }
(\xi -\nu )^{\sigma' }}\right]}.
\end{multline}
$\lambda=\vert m\vert =\vert m' \vert $, $\left\lbrace m,m' \right\rbrace$ are magnetic quantum numbers. See \cite{34_Guseinov_1970, 52_Guseinov_1995} for the explicit form of $g_{\sigma \sigma'}^{q}$. The product of radial parts of NSTOs gives:
\begin{multline}\label{eq:overlap_32}
r_{a}^{n-1}r_{b}^{n'-1}e^{-\zeta_{a}r_{a}-\zeta_{b}r_{b}}\\
=\left(\frac{R}{2}\right)^{n+n'-2}
\left(\xi+\nu\right)^{n-1}
\left(\xi-\nu\right)^{n'-1}
e^{-p\xi-p t \nu},
\end{multline}
This generates the remaining terms, with conditions:

 $n_{2}=n-\sigma > 0$, $n_{3}=n'-\sigma' > 0 $.

First, assuming all the quantum numbers are integers, $\left\lbrace n_{2}, n_{3}\right\rbrace \in \mathbb{Z}$. This defines the two-center overlap integrals for STOs. Thus, the binomial series expansion may be used for the power functions $\left(\xi+\nu \right)^{n_{2}}$, $\left(\xi-\nu \right)^{n_{3}}$;
\begin{multline}\label{eq:overlap_33}
{\mathcal{G}^{n_{1},q}_{n_{2}n_{3}}}(p_{123})
=\frac{p_{1}^{n_{1}}}{\Gamma\left(n_{1}+1\right)}
\sum_{s=0}^{n_{2}+n_{3}}
F_{s}\left(n_{2},n_{3}\right)
\\
\times
\int_{1}^{\infty}\xi^{n_{1}+n_{2}+q-s}e^{-p_{2} \xi}d\xi
\int_{-1}^{1}\nu^{q+s}e^{-p_{3} \nu}d\nu,
\end{multline}
where $F_{s}{\left( n,n' \right)}$: generalized binomial coefficients \cite{34_Guseinov_1970, 51_Pople_1970},
\begin{align}\label{eq:overlap_34}
F_{s}{\left(n,n' \right)}
=\sum_{s'=\frac{1}{2}\left[\left(s-n \right)+\vert s-n \vert \right]}^{min\left(s,n' \right)}(-1)^{s'}F_{s-s'}(n)F_{s'}(n'),
\end{align}
with the coefficients $F_{s}(n)$ are the binomial coefficients indexed by $n$, $s$ which is usually written as $\left(\begin{array}{cc}n\\s\end{array} \right)$ with,
\begin{equation}\label{eq:overlap_35}
\left(\begin{array}{cc}n\\s\end{array} \right)
=\frac{\Gamma\left(n+1\right)}{\Gamma\left(s+1\right)\Gamma\left(n-s+1\right)}.
\end{equation}
If the orbital parameters $\left\lbrace \zeta, \zeta' \right\rbrace$ are equal $\left( p_{2}=p,p_{3}=0 \right)$, the only integral to be computed takes the form:
\begin{equation}\label{eq:overlap_36}
A_{\alpha}\left(p\right)
=\int_{1}^{\infty} \tau^{\alpha}e^{-p \tau}d\tau,
\end{equation}
and for $p>0$ are easily and stably generated by up-ward recursion in $\alpha$ for all positive $p$.\\
If the orbital parameters are different $\left( p_{2}=p,p_{3}=p t \right)$, then an additional and much more difficult integral arises,
\begin{equation}\label{eq:overlap_37}
B_{\alpha}\left(p \right)
=\int_{-1}^{1} \tau^{\alpha}e^{-p \tau}d\tau.
\end{equation}
A down-ward recursive procedure,
\begin{equation}\label{eq:overlap_38}
B_{\alpha}\left(p t \right)
=\frac{1}{p}\left[\alpha B_{\alpha-1}\left(p t \right)+\left(-1\right)^{\alpha}e^{p t}-e^{-p t}  \right],
\end{equation}
for these integrals, stable for all $\alpha$ and $p t$ was given in \cite{53_Cobato_1956}. It uses modified Bessel functions. Bessel functions are first generated by Eq. (\ref{eq:overlap_38}), after which the $B_{\alpha}$ are given as linear combinations of them. This procedure, however, requires more computational effort than an optimal use of up- and down-ward recursion directly in $B_{\alpha}$. By representing the starting values of down-ward recursion formula as incomplete gamma functions, the behavior of the $B_{\alpha}$ integrals was also investigated in \cite{54_Harris_2004}. In another study \cite{55_Guseinov_2007}, in order to calculate the $B_{\alpha}$ integrals for large values of $p t$, the following sum \cite{39_Mulliken_1949} was used:
\begin{equation}\label{eq:overlap_39}
B_{\alpha}\left(p t \right)
=\left(-1\right)^{\alpha+1}A_{\alpha}\left(-p t \right)-A_{\alpha}\left(p t \right).
\end{equation}
For small values of $p t$, the finite series representations of $A_{\alpha}$ integrals,
\begin{equation}\label{eq:overlap_40}
A_{\alpha}\left( p t \right)
=e^{-p t}\sum_{s=1}^{\alpha+1}\frac{\alpha!}{\left(p t\right)^{s}\left(\alpha-s+1 \right)!}
\end{equation}
by replacing $s$ with $\alpha-s+1$; $0 \leq s \leq \alpha$, were used first. Afterwards, the infinite series representations of exponential functions were applied to the resulting expression. Improvements on this formulae were made according to comments in \cite{56_Barnett_2002} on results in \cite{57_Guseinov_1999} (in fact, the formulae given in \cite{58_Guseinov_2002} were more carefully coded rather than any significant change). Regardless of the method, an infinite sum that needs to be accurately calculated, is obtained {\it in fine}.

The only possible simplification in Eq. (\ref{eq:overlap_30}) when the parameters related to quantum numbers take non-integer values $\left\lbrace n_{2}, n_{3}\right\rbrace \in \mathbb{R}$ (Here, we are referring to the two-center overlap integrals for NSTOs),
is eliminating the power functions $\left(\xi\nu\right)^{q}$,
\begin{equation}\label{eq:overlap_41}
\left(\xi \nu\right)^{q}
=\frac{1}{2^{2q}}\sum_{s=0}^{q}\left(-1\right)^{s}F_{s}\left(q\right)
\left(\xi+\nu \right)^{2q-2s} \left(\xi-\nu \right)^{2s}
\end{equation}
The remaining power functions are not analytical which means they can not be represented by a power series \cite{59_Weniger_2008}. Thus,
\begin{equation}\label{eq:overlap_42}
{\mathcal{G}^{n_{1},q}_{n_{2}n_{3}}}(p_{123})
=\frac{1}{2^{2q}}
\sum_{s=0}^{q}\left(-1\right)^{s}F_{s}\left(q \right)
{\mathcal{G}^{n_{1},0}_{n_{2}+2q-2s, n_{3}+2s}}(p_{123}).
\end{equation}
In Eq. (\ref{eq:overlap_42}), $n_{2}$ decreasing while $n_{3}$ increases or visa-versa. Due to the following recurrence relationships, calculating each term arising in Eq. (\ref{eq:overlap_42}) is avoided,\\
down-ward over $n_{2}$, up-ward over $n_{3}$,
\begin{multline}\label{eq:overlap_43}
\left( \frac{p_{2}+p_{3}}{p_{3}} \right){\mathcal{G}^{n_{1},0}_{n_{2}n_{3}}}(p_{123})
\\
=\left( \frac{p_{2}-p_{3}}{p_{3}} \right)\left( \frac{n_{2}}{n_{3}+1} \right)
{\mathcal{G}^{n_{1},0}_{n_{2}-1, n_{3}+1}}(p_{123})
\\
-\frac{p_{2}}{p_{3}}\left(\frac{1}{n_{3}+1}\right)
\Big[\hspace{1mm}
^{+}\mathcal{K}^{n_{1},0}_{n_{2},n_{3}+1}\left(p_{132}\right)
\hspace{0.5mm} - \hspace{0.5mm} ^{+}\mathcal{K}^{n_{1},0}_{n_{3}+1,n_{2}}\left(p_{132}\right)
\Big]
\\
-\left(\frac{1}{n_{3}+1} \right)\mathcal{N}_{n_{2},n_{3}+1}^{n_{1},0}\left(p_{123}\right),
\end{multline}
down-ward over $n_{3}$, up-ward over $n_{2}$,
\begin{multline}\label{eq:overlap_44}
\left( \frac{p_{2}-p_{3}}{p_{3}} \right){\mathcal{G}^{n_{1},0}_{n_{2}n_{3}}}(p_{123})
\\
=-\left( \frac{p_{2}+p_{3}}{p_{3}} \right)\left( \frac{n_{3}}{n_{2}+1} \right)
{\mathcal{G}^{n_{1},0}_{n_{2}+1, n_{3}-1}}(p_{123})
\\
+\frac{p_{2}}{p_{3}}\left(\frac{1}{n_{2}+1}\right)
\Big[\hspace{1mm}
^{+}\mathcal{K}^{n_{1},0}_{n_{3},n_{2}+1}\left(p_{132}\right)
\hspace{0.5mm} - \hspace{0.5mm} ^{+}\mathcal{K}^{n_{1},0}_{n_{2}+1,n_{3}}\left(p_{132}\right)
\Big]
\\
-\left(\frac{1}{n_{2}+1} \right)\mathcal{N}_{n_{2}+1,n_{3}}^{n_{1},0}\left(p_{123}\right),
\end{multline}
with,
\begin{multline}\label{eq:overlap_45}
\mathcal{N}_{n_{2},n_{3}}^{n_{1},q}\left(p_{123}\right)
=\frac{p_{1}^{n_{1}}}{\Gamma\left(n_{1}+1 \right)}e^{-p_{2}}
\\
\times \int_{-1}^{1}\nu^{q}\left(1+\nu \right)^{n_{2}}\left(1-\nu \right)^{n_{3}}e^{-p_{3}\nu}d\nu,
\end{multline}
\begin{multline}\label{eq:overlap_46}
^{+}\mathcal{K}_{n_{2},n_{3}}^{n_{1},q}\left(p_{123}\right)
=\frac{p_{1}^{n_{1}}}{\Gamma\left(n_{1}+1 \right)}e^{-p_{2}}
\\
\times \int_{1}^{\infty} \xi^{q} \left(\xi+1 \right)^{n_{2}}\left(\xi-1 \right)^{n_{3}}e^{-p_{3}\xi}d\xi.
\end{multline}
Note that, the reason for the sign ``$+$'' in definition of the $^{+}\mathcal{K}_{n_{2},n_{3}}^{n_{1},q}$ is that integrals below, in the same form with negative values of $q$ are needed. The integrals are named to emphasize their variables. The constant $\frac{p_{1}^{n_{1}}}{\Gamma\left(n_{1}+1 \right)}$ in Eq. (\ref{eq:overlap_30}) arises from use of Eqs.(\ref{eq:origin_15}, \ref{eq:origin_16}) with Eq. (\ref{eq:origin_19}) according to criterion given in \cite{1_Bagci_2018}. It is thus more advantageous to keep this scheme in sub-functions. See section \ref{k_plus_mines} for explicit forms of the $\mathcal{N}_{n_{2},n_{3}}^{n_{1},q}$ and $^{+}\mathcal{K}_{n_{2},n_{3}}^{n_{1},q}$.\\
Relationships for calculating the starting values of the Eqs. (\ref{eq:overlap_43}, \ref{eq:overlap_44}) are given as:
\begin{multline}\label{eq:overlap_47}
\mathcal{G}^{n_{1},0}_{n_{2}n_{3}}(p_{123})=
\\
\left\lbrace
\begin{array}{ll}
^{\nu}\mathcal{G}^{n_{1},0,0}_{n_{2}n_{3}}(p_{102}) \hspace{28mm}  p_{3}=p t=0
\vspace{2mm}\\
\frac{p_{1}^{n_{1}}}{\Gamma\left(n_{1}+1 \right)}
\sum_{s=0}^{\infty}\left(-1 \right)^{s}\hspace{1mm}{^{\nu}\mathcal{G}^{s,s,s}_{n_{2}n_{3}}(p_{302})}
\hspace{2mm} p_{3}=p t \neq 0
\end{array}
\right.
\end{multline}
where,
\begin{multline}\label{eq:overlap_48}
{^{\nu}\mathcal{G}^{n_{1},q_{1},q_{2}}_{n_{2}n_{3}}(p_{123})}
=\frac{p_{1}^{n_{1}}}{\Gamma\left(n_{1}+1 \right)}e^{-p_{2}}
\\
\times \int_{1}^{\infty}\int_{-1}^{1}\xi^{-q_{1}}\left(\xi \nu \right)^{q2}
\left(\xi+\nu \right)^{n_{2}}\left(\xi-\nu \right)^{n_{3}}e^{-p_{3}\xi}d\xi d\nu .
\end{multline}
Please see Appendices (\ref{sec:appendicesa}, \ref{sec:appendicesb}) for derivations of the Eqs. (\ref{eq:overlap_43}, \ref{eq:overlap_44}) and Eq. (\ref{eq:overlap_47}), respectively.\\
Using Eq. (\ref{eq:overlap_41}) again, we have,
\begin{multline}\label{eq:overlap_49}
{^{\nu}\mathcal{G}^{n_{1},q_{1},q_{2}}_{n_{2}n_{3}}(p_{123})}
\\
=\frac{1}{2^{2 q_{2}}}
\sum_{s=0}^{q_{2}}\left(-1\right)^{s}F_{s}\left(q_{2} \right)
\hspace{1mm}
{^{\nu}\mathcal{G}^{n_{1},q_{1},0}_{n_{2}+2q_{2}-2s,n_{3}+2s}(p_{123})}.
\end{multline}
The right hand-side of  Eq. (\ref{eq:overlap_49}) was previously expressed in terms of incomplete beta functions \cite{1_Bagci_2018} as,
\begin{multline}\label{eq:overlap_50}
{^{\nu}\mathcal{G}^{n_{1},q_{1},0}_{n_{2}n_{3}}(p_{123})}
=2^{q_{1}}\left[ ^{1}\mathcal{K}^{n_{1},q_{1}}_{n_{2}n_{3}}\left(p_{123}\right)
\hspace{1mm}+\hspace{1mm} ^{1}\mathcal{K}^{n_{1},q_{1}}_{n_{3}n_{2}}\left(p_{123}\right)\right.
 \\
-^{2}\mathcal{K}^{n_{1},q_{1}}_{n_{2}n_{3}}\left(p_{123}\right)
\left. \hspace{1mm}-\hspace{1mm} ^{2}\mathcal{K}^{n_{1},q_{1}}_{n_{3}n_{2}}\left(p_{123}\right) \right],
\end{multline}
where,
\begin{multline}\label{eq:overlap_51}
^{1}\mathcal{K}^{n_{1},q}_{n_{2}n_{3}}\left(p_{123}\right)
=\frac{p_{1}^{n_{1}}}{\Gamma\left(n_{1}+1 \right)}e^{-p_{2}}
\\
\times \int_{1}^{\infty}\left(2\xi \right)^{n_{2}+n_{3}-q+1}B_{n_{2}+1,n_{3}+1}\left(\frac{\xi+1}{2\xi} \right) e^{-p_{3}\xi}d\xi,
\end{multline}
\begin{multline}\label{eq:overlap_52}
^{2}\mathcal{K}^{n_{1},q}_{n_{2}n_{3}}\left(p_{123}\right)
=\frac{p_{1}^{n_{1}}}{\Gamma\left(n_{1}+1 \right)}e^{-p_{2}}
\\
\times \int_{1}^{\infty}\left(2\xi \right)^{n_{2}+n_{3}-q+1}B_{n_{2}+1,n_{3}+1}\left(\frac{1}{2} \right)e^{-p_{3}\xi}d\xi,
\end{multline}
with incomplete beta functions,
\begin{equation}\label{eq:overlap_53}
B_{nn'}\left(z \right)
=\int_{0}^{z}\tau^{n-1}(1-\tau)^{n'-1}d\tau,
\end{equation}
These may be found in:  \cite{36_Abramowitz_1972}. Analytical relations were derived using the following identity in  Eqs. (\ref{eq:overlap_49}),
\begin{equation}\label{eq:overlap_54}
B_{nn'}\left(1-z\right)=B_{nn'}-B_{nn'}\left(z \right),
\end{equation}
$B_{nn'}$ are beta functions, and using their series representation \cite{60_Temme_2011} of the incomplete beta functions,
\begin{equation}\label{eq:overlap_55}
B_{nn'}\left(z\right)
=\sum_{s=0}^{\infty}\frac{\left(1-n'\right)_{s}}{\left(n+s \right)s!}z^{n+s}; \hspace{3mm} \vert z\vert <1.
\end{equation}
Considering the domains of integrals given in Eq. (\ref{eq:overlap_30}) and the Eq. (\ref{eq:overlap_54}), the convergence condition: $1-z=\frac{\xi-1}{2\xi}<\frac{1}{2}$, is satisfied. Thus, there are no significant computational disadvantages of using non-integer principal quantum numbers in Slater-type orbitals while the orbital parameters are equal. Otherwise, it is necessary to use the series representation of exponential functions in addition, where the upper limit of summation increases depending on values of $p_{3}\left( p t \right)$ (for large values, precise results require more terms). If, however, the summations arising from series representation of exponential functions are also transformed into appropriate recurrence relations then the number of terms to be used in the summation is no longer a drawback.

Skipping the procedure given between Eqs. (\ref{eq:overlap_51}-\ref{eq:overlap_55}) and continuing directly from the Eq. (\ref{eq:overlap_50}), e.g. using recurrence relations \cite{60_Temme_2011} for incomplete beta functions and integration by parts gives recursion as follows: \\
down-ward over $n_{2}$, up-ward over $n_{3}$,
\begin{multline}\label{eq:overlap_56}
{^{\nu}\mathcal{G}^{n_{1},q_{1},0}_{n_{2}n_{3}}(p_{123})}
=\frac{B_{n_{2}+1,n_{3}+1}}{B_{n_{3}+2,n_{2}}}\hspace{1mm} {^{\nu}\mathcal{G}^{n_{1},q_{1},0}_{n_{2}-1,n_{3}+1}(p_{123})}
\\
+\left(\frac{1}{n_{3}+1}\right)
\big[\hspace{1mm}
^{-}\mathcal{K}_{n_{3}+1,n_{2}}^{n_{1},q_{1}}\left(p_{123}\right)
\hspace{0.5mm} - \hspace{0.5mm}^{-}\mathcal{K}_{n_{2},n_{3}+1}^{n_{1},q_{1}}\left(p_{123}\right)
\big],
\end{multline}
down-ward over $n_{3}$, up-ward over $n_{2}$,
\begin{multline}\label{eq:overlap_57}
{^{\nu}\mathcal{G}^{n_{1},q_{1},0}_{n_{2}n_{3}}(p_{123})}
=\frac{B_{n_{2}+1,n_{3}+1}}{B_{n_{2}+2,n_{3}}}\hspace{1mm} {^{\nu}\mathcal{G}^{n_{1},q_{1},0}_{n_{2}+1,n_{3}-1}(p_{123})}
\\
+\left(\frac{1}{n_{2}+1}\right)
\big[\hspace{1mm}
^{-}\mathcal{K}_{n_{2}+1,n_{3}}^{n_{1},q_{1}}\left(p_{123}\right)
\hspace{0.5mm} - \hspace{0.5mm} ^{-}\mathcal{K}_{n_{3},n_{2}+1}^{n_{1},q_{1}}\left(p_{123}\right)
\big],
\end{multline}
here,
\begin{multline*}
\big[\hspace{1mm}
^{-}\mathcal{K}_{n_{2}+1,n_{3}}^{n_{1},q_{1}}\left(p_{123}\right)
\hspace{0.5mm} - \hspace{0.5mm} ^{-}\mathcal{K}_{n_{3},n_{2}+1}^{n_{1},q_{1}}\left(p_{123}\right)
\big]
\\
=\big[\hspace{1mm}
^{-}\mathcal{K}_{n_{2},n_{3}}^{n_{1},q_{1}+1}\left(p_{123}\right)
\hspace{0.5mm} - \hspace{0.5mm} ^{-}\mathcal{K}_{n_{3},n_{2}}^{n_{1},q_{1}+1}\left(p_{123}\right)
\big],
\end{multline*}
$^{-}\mathcal{K}_{n_{2},n_{3}}^{n_{1},q} \hspace{0.5mm} = \hspace{0.5mm} ^{+}\mathcal{K}_{n_{2},n_{3}}^{n_{1},-q}$; $^{-}\mathcal{K}_{n_{2},n_{3}}^{n_{1},0} \hspace{0.5mm} = \hspace{0.5mm} ^{+}\mathcal{K}_{n_{2},n_{3}}^{n_{1},0}$.
\vskip3mm
The Eqs. (\ref{eq:overlap_42}, \ref{eq:overlap_49}) are similar. They run over fixed values of $\left\lbrace q,q_{2} \right\rbrace$. On the other hand, the parameter $q_{1}$ is dynamic due to the Eq. (\ref{eq:overlap_47}). Using  Eq. (\ref{eq:overlap_50}) again but this time using derivatives of the incomplete beta functions, the recurrence relations over $q_{1}$ are obtained as,
\begin{multline}\label{eq:overlap_58}
{^{\nu}\mathcal{G}^{n_{1},q_{1},0}_{n_{2}n_{3}}(p_{123})}
=\left( \frac{1}{n_{2}+n_{3}-q_{1}+2} \right)
\big[p_{3}\hspace{0.5mm}
{^{\nu}\mathcal{G}^{n_{1},q_{1}-1,0}_{n_{2}n_{3}}(p_{123})}
\big.
\\
+\hspace{0.5mm} ^{-}\mathcal{K}_{n_{2},n_{3}}^{n_{1},q_{1}}\left(p_{123}\right)
+\hspace{0.5mm} ^{-}\mathcal{K}_{n_{3},n_{2}}^{n_{1},q_{1}}\left(p_{123}\right)
\big].
\end{multline}
Finally, for starting values of $^{-}\mathcal{K}_{n_{2},n_{3}}^{n_{1},q_{1}}\left(p_{123}\right)$, we have,
\begin{multline}\label{eq:overlap_59}
^{1}\mathcal{K}_{n_{2},n_{3}}^{n_{1},q_{1}}\left(p_{123}\right)
=\frac{B_{n_{2}+1,n_{3}+2}}{B_{n_{2}+2,n_{3}}}
\hspace{1mm}^{1}\mathcal{K}_{n_{2}+1,n_{3}-1}^{n_{1},q_{1}}\left(p_{123}\right)
\\
+\frac{1}{2^q\left(n_{2}+1 \right)}
\hspace{0.5mm}
^{-}\mathcal{K}_{n_{2},n_{3}}^{n_{1},q_{1}}\left(p_{123}\right).
\end{multline}
The used recurrence relations and derivatives for incomplete beta functions are given in the Appendix \ref{sec:appendicesc}.
\section{Relationships for $^{+}\mathcal{K}_{n_{2},n_{3}}^{n_{1},q_{1}}$ and $^{-}\mathcal{K}_{n_{2},n_{3}}^{n_{1},q_{1}}$ integrals}\label{k_plus_mines}
In order to obtain recurrence relations for $^{+}\mathcal{K}_{n_{2},n_{3}}^{n_{1},q_{1}}$ and $^{-}\mathcal{K}_{n_{2},n_{3}}^{n_{1},q_{1}}$ integrals, first, from the following identities of $\left(\frac{1}{\xi^{q_{1}}} \right)$ it is easy to write,
\begin{equation*}
\frac{1}{\xi^{q_{1}}}
=\left(\frac{\xi+1}{\xi} -1\right)^{q_{1}}
=\left(1-\frac{\xi-1}{\xi}\right)^{q_{1}}
=\left(\xi-\frac{\xi^2-1}{\xi} \right)^{q_{1}},
\end{equation*}
Thus,
\begin{subequations}\label{eq:k_plus_mines_60}
\begin{equation}\label{eq:k_plus_mines_60a}
^{-}\mathcal{K}_{n_{2},n_{3}}^{n_{1},q_{1}}
=
\sum_{s=0}^{q_{1}}\left(-1 \right)^{q_{1}-s}F_{s}\left(q_{1} \right)
\hspace{0.5mm}
^{-}\mathcal{K}_{n_{2}+s,n_{3}}^{n_{1},s},
\end{equation}
\begin{equation}\label{eq:k_plus_mines_60b}
^{-}\mathcal{K}_{n_{2},n_{3}}^{n_{1},q_{1}}
=
\sum_{s=0}^{q_{1}}\left(-1 \right)^{s}F_{s}\left(q_{1} \right)
\hspace{0.5mm}
^{-}\mathcal{K}_{n_{2},n_{3}+s}^{n_{1},s},
\end{equation}
\begin{multline}\label{eq:k_plus_mines_60c}
^{-}\mathcal{K}_{n_{2},n_{3}}^{n_{1},q_{1}}
=
\sum_{s=0}^{\lfloor q_{1}/2 \rfloor}\left(-1 \right)^{s}F_{s}\left(q_{1} \right)
\hspace{0.5mm}
^{+}\mathcal{K}_{n_{2}+s,n_{3}+s}^{n_{1},s}
\\
+\sum_{s=\lfloor q_{1}/2 \rfloor +1}^{q_{1}}\left(-1 \right)^{s}F_{s}\left(q_{1} \right)
\hspace{0.5mm}
^{-}\mathcal{K}_{n_{2}+s,n_{3}+s}^{n_{1},s}.
\end{multline}
\end{subequations}
The calculations start with Eq. (\ref{eq:k_plus_mines_60c}). The first summation on the right hand-side is reduced over $s$ as follows,
\begin{equation}\label{eq:k_plus_mines_61}
{^{+}\mathcal{K}^{n_{1},s}_{n_{2}n_{3}}}
=\frac{1}{2^{2 s}}
\sum_{s'=0}^{s}\left(-1\right)^{s'}F_{s'}\left(s \right)
\hspace{0.5mm}
{^{+}\mathcal{K}^{n_{1},0}_{n_{2}+2s-2s',n_{3}+2s'}}.
\end{equation}
Considering the identities $\xi=\left(\xi+1\right)-1$, $\xi=\left(\xi-1\right)+1$, we also have,
\begin{subequations}\label{eq:k_plus_mines_62}
\begin{equation}\label{eq:k_plus_mines_62a}
^{+}\mathcal{K}_{n_{2},n_{3}}^{n_{1},s}
=
\sum_{s'=0}^{s}\left(-1 \right)^{s-s'}F_{s'}\left(s \right)
\hspace{0.5mm}
^{+}\mathcal{K}_{n_{2}+s',n_{3}}^{n_{1},0}
\end{equation}
\begin{equation}\label{eq:k_plus_mines_62b}
^{+}\mathcal{K}_{n_{2},n_{3}}^{n_{1},s}
=
\sum_{s'=0}^{s}\left(-1 \right)^{s'}F_{s'}\left(s \right)
\hspace{0.5mm}
^{+}\mathcal{K}_{n_{2},n_{3}+s'}^{n_{1},0}.
\end{equation}
\end{subequations}
For the starting values we can use the following relationship,
\begin{multline}\label{eq:overlap_63}
\displaystyle
^{+}\mathcal{K}_{n_{2},n_{3}}^{n_{1},q}\left(p_{123}\right)
=\frac{p_{1}^{n_{1}}}{\Gamma\left(n_{1}+1 \right)}e^{-p_{3}-p_{2}}
\\
\times \sum_{s=0}^{q}F_{s}\left(q \right)
\Bigg\lbrace
\left(-1\right)^{q+s}2^{n_{2}+n_{3}+s+1}
\frac{\Gamma\left(-n_{2}-n_{3}-q-1 \right)}{\Gamma\left(-n_{2} \right)}
\\
\times \frac{\Gamma\left(n_{2}+n_{3}+q+2\right)}{\Gamma\left(n_{2}+n_{3}+s+2 \right)}
{_{1}}F_{1}\left(n_{3}+s+1;n_{2}+n_{3}+s+2;2p_{2} \right)
\\
+ \frac{\Gamma\left(n_{2}+n_{3}+q-s+1\right)}{p_{2}^{n_{2}+n_{3}+q-s+1}}
{_{1}}F_{1}\left(-n_{2};-n_{2}-n_{3}-q+s;2p_{2} \right)
\Bigg\rbrace,
\\
\left\lbrace n_{2}, n_{3}, n_{2}+n_{3} \right\rbrace \notin \mathbb{N}; \hspace{1mm} q \geq 0,
\end{multline}
which has quite simple form while $q=0.$\\
The second summation of the Eq. (\ref{eq:k_plus_mines_60c}) to reduce over $s$, the recurrence relation obtained via integration by parts is used,
\begin{multline}\label{eq:k_plus_mines_64}
s \hspace{0.5mm} ^{-}\mathcal{K}_{n_{2},n_{3}}^{n_{1},s+1}
=\left(2^{n_{2}}\delta_{n_{3},0}+2^{n_{3}}\delta_{n_{2},0} \right)e^{-p_{3}}
-p_{3}\hspace{0.5mm} ^{-}\mathcal{K}_{n_{2},n_{3}}^{n_{1},s}
\\
+\left(n_{2}+n_{3}\right)\hspace{0.5mm} ^{-}\mathcal{K}_{n_{2}-1,n_{3}-1}^{n_{1},s-1}
-\left(n_{2}-n_{3} \right)\hspace{0.5mm} ^{-}\mathcal{K}_{n_{2}-1,n_{3}-1}^{n_{1},s}.
\end{multline}
At the end of this process, all the terms are in form that $^{-}\mathcal{K}_{n_{2}+s,n_{3}+s}^{n_{1},1}$. Now, using Eqs. (\ref{eq:k_plus_mines_60a}, \ref{eq:k_plus_mines_60b}) while $s=1$,
\begin{subequations}\label{eq:k_plus_mines_65}
\begin{equation}\label{eq:k_plus_mines_65a}
^{-}\mathcal{K}_{n_{2}+s-1,n_{3}}^{n_{1},1}
=
\hspace{0.5mm}
^{-}\mathcal{K}_{n_{2}+s,n_{3}+s}^{n_{1},1}
-\hspace{0.5mm}
^{+}\mathcal{K}_{n_{2}+s-1,n_{3}}^{n_{1},0},
\end{equation}
\begin{equation}\label{eq:k_plus_mines_65b}
^{-}\mathcal{K}_{n_{2},n_{3}+s-1}^{n_{1},1}
=
\hspace{0.5mm}
^{+}\mathcal{K}_{n_{2}+s-1,n_{3}}^{n_{1},0}
-\hspace{0.5mm}
^{-}\mathcal{K}_{n_{2}+s,n_{3}+s}^{n_{1},1}.
\end{equation}
\end{subequations}
Finally all the terms are expressed in terms of $^{-}\mathcal{K}_{n_{2},n_{3}}^{n_{1},1}$ which is analytically calculated by Eq. (\ref{eq:overlap_59}) based on \cite{1_Bagci_2018}. Note that, similarly relationships for $\mathcal{N}_{n_{2},n_{3}}^{n_{1},q}$ are derived. Here we give only an expression required to calculate the starting values,
\begin{multline}\label{eq:k_plus_mines_66}
\mathcal{N}_{n_{2},n_{3}}^{n_{1},q}\left(p_{123}\right)
=\frac{p^{n_{1}}}{\Gamma\left(n_{1}+1 \right)}
2^{n_{2}+n_{3}+1}e^{p_{3}-p_{2}}
\Gamma\left(n_{3}+1 \right)
\\
\times \sum_{s=0}^{q}\left(-1 \right)^{q+s}2^{s}F_{s}\left(q \right)
\frac{\Gamma\left(n_{2}+s+1 \right)}{\Gamma\left(n_{2}+n_{3}+s+2 \right)}
\\
\times {_{1}}F_{1}\left(n_{2}+s+1;n_{2}+n_{3}+s+2;-2p_{3} \right).
\end{multline}

{\bf \large Summary of the integral method:}

The procedure of calculation has the following order,
\begin{itemize}
\item{Eq.(\ref{eq:overlap_59}) through Eq.(\ref{eq:overlap_51}) [See \cite{1_Bagci_2018} or use Eqs.(\ref{eq:overlap_54}-\ref{eq:overlap_55})] is calculated for $q_{1}=1.$}

\item{The results are used in the Eq. (\ref{eq:overlap_58}), where the first term on the right hand-side corresponds to $p t=0$ in Eq.(\ref{eq:overlap_47}).}

\item{The value of $q_{1}$ is increased. Eqs. (\ref{eq:k_plus_mines_60}-\ref{eq:k_plus_mines_65}) are used to calculate $^{-}\mathcal{K}_{n_{3},n_{2}}^{n_{1},q_{1}}$ for that value.}

\item{The Eq.(\ref{eq:overlap_56}) or Eq. (\ref{eq:overlap_57}) is used. They give all the terms arising in the Eq. (\ref{eq:overlap_49}) and Eq. (\ref{eq:overlap_47}).}

\end{itemize}

\section{Conclusion}\label{sec:conclusion}
Until few years ago, the use of Slater basis sets with non-integer principal quantum numbers in electron structure calculation was thought to be nearly impossible \cite{59_Weniger_2008}. The attempts in this regard failed due to the absence of benchmark values. The authors in their previous work \cite{2_Bagci_2015, 61_Bagci_2014, 62_Bagci_2015} first focused on this issue e.g., obtaining results for molecular integrals with unquestionable precision via numerical techniques. They presented benchmark values for two- and three-center integrals. The non-integer Slater-type orbitals are not analytic at $r=0$ in the sense of complex analysis. From the mathematical point of view this implies that analytically closed form relations for molecular integrals are unavailable.  Compact form relations, however, avoid using series representation of incomplete gamma functions which are not stable for all values of parameters. They were established in the first paper of this series \cite{1_Bagci_2018}. Analytical relations for the first term $\left(p\tau=0 \right)$ of Eq. (\ref{eq:overlap_47}) were derived in terms of incomplete beta functions. The previous paper also allows us to argue the increased efficiency of the integration procedure outlined above. Note that an approximate count of FLOPs needed is in favour of the strategy outlined in the present paper. As emphasized in closing this section, the next step is to develop a suitable code for rapid numerical evaluation of the integrals according to the method developed here. In the present study, they are used as a starting point to calculate the second term $\left(p\tau \neq 0 \right)$ via recurrence relations. This is similar to a method used for calculation of overlap-like integrals while $\left(n_{2}, n_{3} \right) \in \mathbb{Z}$ [see the Eq. (\ref{eq:overlap_39})] and leads calculating series representation of incomplete beta functions only once and for all. The relationships obtained are expressed in terms of $^{-}\mathcal{K}_{n_{3},n_{2}}^{n_{1},q_{1}}$. Accordingly, the related section (\ref{k_plus_mines}) contains an explicit description.

The procedure of integral calculation, outlined above generates a stable algorithm. This permits efficient calculation of the molecular integrals. Developing a computer program for two-center, one- and two-electron integrals by using the formulae given here and obtaining results for different orbital parameters and inter-nuclear distances will be the subject of next research.

\begin{appendices}
\section*{Appendices}\label{sec:appendices}
\subsection{}\label{sec:appendicesa}
Applying the integration by parts twice to the Eq. (\ref{eq:overlap_30}) while $q=0$, where in the first operation we use,\\
for the Eq. (\ref{eq:overlap_43}),
\begin{equation*}
U=\left(\xi+\nu \right)^{n_{2}}e^{-p_{2}\xi-p_{3}\nu},
\hspace{2mm}
dV=\left(\xi-\nu \right)^{n_{3}}d\xi,
\end{equation*}
for the Eq. (\ref{eq:overlap_44}),
\begin{equation*}
U=\left(\xi-\nu \right)^{n_{3}}e^{-p_{2}\xi-p_{3}\nu},
\hspace{2mm}
dV=\left(\xi+\nu \right)^{n_{2}}d\xi,
\end{equation*}
thus we have,\\
\begin{multline}\label{eq:appendicesa_67}
\mathcal{G}^{n_{1},0}_{n_{2}n_{3}}(p_{123})
=\frac{p_{2}}{\left(n_{3}+1 \right)}
\mathcal{G}^{n_{1},0}_{n_{2},n_{3}+1}(p_{123})
\\
-\frac{n_{2}}{\left(n_{3}+1 \right)}
\mathcal{G}^{n_{1},0}_{n_{2}-1,n_{3}+1}(p_{123})
-\frac{1}{\left(n_{3}+1 \right)}
\mathcal{N}^{n_{1},0}_{n_{2},n_{3}+1}(p_{123}),
\end{multline}
\begin{multline}\label{eq:appendicesa_68}
\mathcal{G}^{n_{1},0}_{n_{2}n_{3}}(p_{123})
=\frac{p_{2}}{\left(n_{2}+1 \right)}
\mathcal{G}^{n_{1},0}_{n_{2}+1,n_{3}}(p_{123})
\\
-\frac{n_{3}}{\left(n_{2}+1 \right)}
\mathcal{G}^{n_{1},0}_{n_{2}+1,n_{3}-1}(p_{123})
-\frac{1}{\left(n_{2}+1 \right)}
\mathcal{N}^{n_{1},0}_{n_{2}+1,n_{3}}(p_{123}).
\end{multline}
Performing the second operation to the first terms on the right-hand side of Eqs. (\ref{eq:appendicesa_67}, \ref{eq:appendicesa_68}) with,
\begin{equation*}
U=\left(\xi+\nu \right)^{n_{2}}\left(\xi-\nu \right)^{n_{3}+1},
\hspace{2mm}
dV=e^{-p_{3}\nu}d\nu,
\end{equation*}
\begin{equation*}
U=\left(\xi+\nu \right)^{n_{2}+1}\left(\xi-\nu \right)^{n_{3}},
\hspace{2mm}
dV=e^{-p_{3}\nu}d\nu,
\end{equation*}
the Eqs. (\ref{eq:overlap_43}, \ref{eq:overlap_44}) are obtained, respectively.
\subsection{}\label{sec:appendicesb}
Using the series expansion of exponential functions $e^z$, where $z=-p_{3}\nu$ in the Eq. (\ref{eq:overlap_30}) while $q=0$ we have,
\begin{multline}\label{eq:appendicesb_69}
\mathcal{G}^{n_{1},0}_{n_{2}n_{3}}(p_{123})
=\frac{p_{1}^{n_{1}}}{\Gamma\left(n_{1}+1 \right)}
\sum_{s=0}^{\infty}\left(-1 \right)^{s}\frac{p_{3}^{s}}{\Gamma\left(s+1 \right)}
\\
\times \int_{1}^{\infty}\int_{-1}^{1}
\nu^{s}\left(\xi+\nu \right)^{n_{2}}\left(\xi-\nu \right)^{n_{3}}e^{-p_{3}\xi}d\xi d\nu.
\end{multline}
Dividing and multiplying the integral on the right-hand side with $\xi^{s}$ gives:
\begin{multline}\label{eq:appendicesb_70}
{^{\nu}\mathcal{G}^{n_{1},s,s}_{n_{2}n_{3}}(p_{123})}
=\frac{p_{3}^{s}}{\Gamma\left(s+1 \right)}
\\
=\int_{1}^{\infty}\int_{-1}^{1}
\xi^{-s} \left(\xi \nu \right)^{s}
\left(\xi+\nu\right)^{n_{2}}
\left(\xi-\nu\right)^{n_{3}}
e^{-p_{2}\xi} d\xi d\nu.
\end{multline}
The Eq. (\ref{eq:overlap_48}) is obtained by analogously generalization of power functions $\xi^{-s}$, $\left(\xi \nu \right)^{s}$ to $\xi^{-q_{1}}$, $\left(\xi \nu \right)^{q_{2}}$, respectively.
\subsection{}\label{sec:appendicesc}
The normalized version of incomplete beta functions are used,
\begin{equation}\label{eq:appendicesc_71}
\mathfrak{B}_{nn'}\left(z \right)=
\frac{B_{nn'}\left(z \right)}{B_{nn'}};
\hspace{2mm}
\mathfrak{B}_{nn'}\left(z \right)
=1-\mathfrak{B}_{n'n}\left(1-z \right).
\end{equation}
They are usually represented by ``$I$'' hovewer we use ``$\mathfrak{B}$'' in order to avoid the confusion with the Bessel functions. Thus, the recurrence relations and the derivatives used in the present work have the following form,
\begin{subequations}\label{eq:appendicesc_72}
\begin{equation}\label{eq:appendicesc_72a}
\mathfrak{B}_{nn'}\left(z \right)
=\mathfrak{B}_{n+1,n'-1}\left(z \right)
+\frac{z^{n}\left(1-z\right)^{n'-1}}{nB_{nn'}},
\end{equation}
\begin{equation}\label{eq:appendicesc_72b}
\mathfrak{B}_{nn'}\left(z \right)
=\mathfrak{B}_{n-1,n'+1}\left(z \right)
+\frac{z^{n-1}\left(1-z\right)^{n'}}{n'B_{n'n}},
\end{equation}
\begin{equation}\label{eq:appendicesc_73c}
\frac{\partial \mathfrak{B}_{nn'}\left(z \right)}{\partial z}
=\frac{\left(z\right)^{n-1}\left(1-z\right)^{n'-1}}{B_{nn'}},
\end{equation}
\end{subequations}
$B_{nn'}=B_{n'n}$ and $z=\frac{\xi+1}{2\xi}$, $1-z=\frac{\xi-1}{2\xi}$.

\end{appendices}

\end{document}